\documentclass{ws-ijmpb}

\begin{document}

\markboth{J. R\"ohler}
{The bulge in the basal plane}

%
%

\title{THE BULGE IN THE BASAL PLANE AREA OF CUPRATE SUPERCONDUCTORS }

\author{J. R\"ohler}

\address{Universit\"at zu K\"oln, Fachgruppe Physik, Z\"ulpicher Str. 
77, D-50937 K\"oln, Germany \\
 abb12@uni-koeln.de}

\maketitle

\begin{history}
\received{11 July 2004}
\end{history}

\begin{abstract}
The bulge in the doping dependence of the basal plane area of cuprate 
superconductors is shown to be an effect of the particular inhomogenous 
electronic structure created by the dense packing of paired self-protected 
singlets (PSPS) in CuO$_2$ lattices.
\end{abstract}

\keywords{superconductivity; cuprates; electronic structure; lattice}

\section{The experimental evidence}	%
Hole doping of cuprate superconductors removes electrons from the
antibonding planar Cu$3d_{x^2-y^2}$O$2p_{xy}$ states; thus increasing
doping is expected to shorten the planar Cu--O bonds.  All accessible
crystallographic data, however, give evidence for a significant
deviation from the Pauling-type contraction of covalent bonds on 
increasing covalency. 
Between the onset of superconductivity at $n_{I-M}=0.07
$(2), and the closure of the large pseudogap at $n_{PG}=0.21$(2) the
basal plane area ($a^2$ or $ab$) is always found \cite{1} concave away from the
$n_{h}$ axis  instead of convex toward it. This ``bulge''
exhibits a maximum around optimum doping $n_{opt}=0.16$(1), and
collapses at $n_{PG}=0.21$(2), notably $within$ the overdoped regime. 
Optimum doping may inflate the basal plane area by up to 30\% of its 
overall variation.

\section{The model}
\subsection{Inhomogenous electronic structure}
Apparently hole concentrations $n_{h}\leq n_{opt}\simeq 0.16$ exert an 
anomalous {\it outward} electronic pressure on the basal Cu grid 
counteracting the normal compressive covalent strain. 
%
%
\begin{figure}[th]
\centerline{\psfig{file=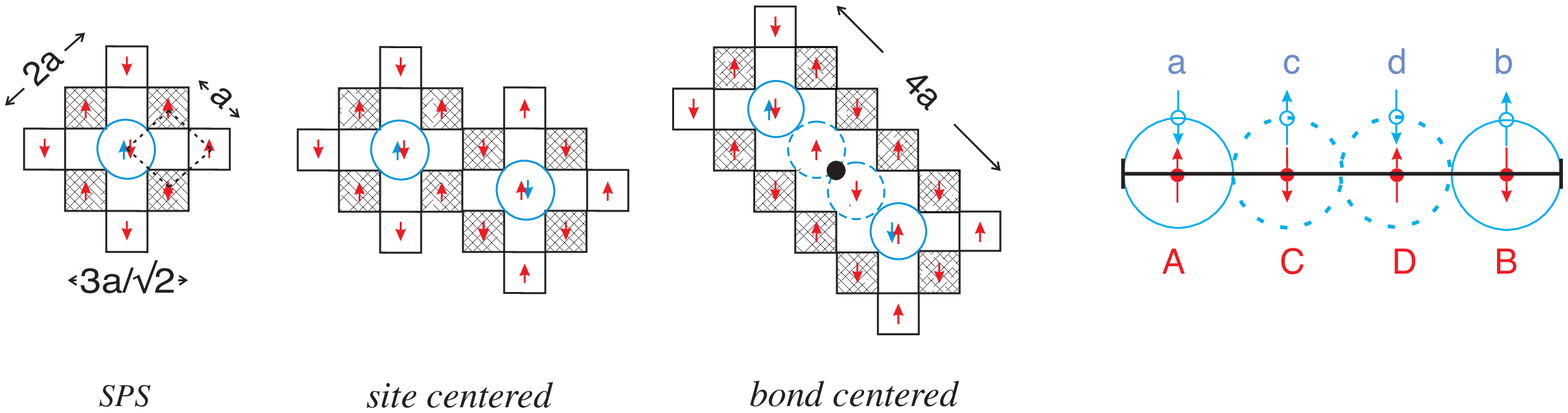,width=12.5cm,trim=0 8cm 2cm 
5cm}}
\vspace*{8pt}
\caption{$Left$: Self-protecting singlet (SPS) in its 
antiferromagnetic environment, and their two possible 
$nn$ configurations. Circles: cages of ZR singlets; hatched squares : excluded 
areas. $Right$: Schematized internal structure of a paired 
self-protecting singlet (PSPS). Closed circles indicate the two paired 
ZR singlets, dashed circles intermediate ZR singlets.) }
\end{figure}
We connect the anomalous outward electronic pressure with the $Aufbau$
principles of the many body state in hole-doped square-planar
CuO$_{2}$ lattices.  On the high energy scale of Coulomb ($U$) and
exchange ($J$) interactions there is little doubt that the
non-double-occupancy constraint for holes at the Cu$3d^9$ sites
($U_{dd}\gg t$) is the basic rule that determines the electronic
structure.  It is a necessary, however, not a sufficient rule.  Here
we propose that the characteristic hole concentrations of the bulge,
0.07(2), 0.16(1), and 0.21(2), point to the effect of an additional
non-double-occupancy constraint for doped holes in a corner-linked,
square CuO$_2$ lattice.  Therein corner linked ``cages'' of 4 oxygens
(squares in Fig.  1) enclose single Cu sites.  Due to the phase
coherence in the symmetric combination of the four O$2p_{x,y}$
orbitals, holes doped into the cage form extraordinarily stable
spin-singlets with the central Cu spins. These ``Zhang-Rice singlets'' 
(ZR) do not allow for an occupancy of their cages with two oxygen holes. 
The repulsion between two nonorthogonal ZR singlets
may be estimated as $R_{nn}\sim U_{pp}/32\simeq 0.08$ eV.
$U_{pp}=2.6$ eV is the oxygen on-site Coulomb repulsion.  The 
effective repulsion between overlapping ZR singlets will create an excluded area 
extending over the four $nn$ cages and ``protect'' the spin-singlet 
in the central cage\cite{2}.  A scheme for such a ``self-protecting singlet'' (SPS) 
is given in Fig. 1 (left). Hole doping obeying $both$ 
non-double-occupancy constraints will populate the CuO$_2$ lattice with SPS, 
instead of overlapping  ZR or simple RVB singlets. 
Two neigboring SPS may connect each other 
in a ``site centered'', or  a ``bond centered'' configuration, (see Fig. 1, middle). 
The latter has consequences for pairing, since its connecting site 
(black dot) is located at the $a$\/-axis, and is an inversion center for all 
space and spin coordinates. Two SPS connected along the Cu--O bonds 
thus may in principle exchange their oxygen holes, and form 
a ``paired self-protecting singlet'' (PSPS).

Copper and oxygen hole states in a PSPS are not dynamically
independent from each other: a spin-singlet of the oxygen holes 
in cages a and b (Fig. 1, right) will be strongly correlated with the 
antiferromagnetic chain of Cu spins, ACDB, that polarizes the delocalized oxygen 
hole spins. In addition a PSPS is expected to develop a boson-fermion structure. 
For example, a spin-singlet of the fermionic oxygen holes in cages a 
and b will glue together the two bosonic ZR singlets, and thus form a strongly
anisotropic bosonic pair. Moreover, any attractive exchange 
interaction between cages a and b has to proceed via two repulsive 
intermediate triplet excitations, repelling the sites 
(A--a) from (D--d), and (B--b) from (C--c). Thus two bosonic ZR
singlets bound by a singlet of their delocalized fermionic 
oxygen holes will effectively $expand$ the PSPS along the Cu--O direction.

Doping of the CuO$_2$ lattice with intact PSPS and SPS is 
geometrically limited by their closest packings: 
$n_{PSPS}=1/6=0.166\simeq n_{opt}=0.16(1)$, and $n_{SPS}=1/5=0.2$. 
Overdoping with $n_{h}>n_{PSPS}=0.166$ will start
destroying the inversion center of the PSPS and therewith the 
condition for pairing. Then at $n_{h}>n_{SPS}=0.2$ 
the oxygen cages will also start to be destroyed. 
At $n_{h}\geq 2/9=0.22$ all oxygen cages 
will be broken and ZR singlets may not form any longer. Consequently 
$R_{nn}\rightarrow 0$ at $n_{h}= 0.22\simeq n_{PG}$. It is thus 
appropriate to connect optimum doping with a many body state of the 
densest packed PSPS, and the closure of the (large) pseudogap with the 
destruction of all bosonic ZR singlets. As a consequence the electronic structure 
up to optimum doping is expected to be inhomogenous, 
in the sense that the charge, spin and lattice dynamics are determined 
by an intrinsic pairing length of $4a$ along the Cu--O bond 
directions. An adiabatic snapshot of the electronic structure 
taken at $n_{opt}$ would capture a ``tweed-like'' pattern 
of densest packed PSPS zig-zagging  along the Cu--O directions (Fig. 2, left).

\subsection{Lattice dynamics and low-$T$ charge ordering}
How may the two oxygen holes in a PSPS achieve the necessary overlap
for pairing by exchange?  Perfectly rigid square lattices will
certainly tend trapping the holes in their oxygen cages. The lattice dynamics of 
doped pervoskites is, however, disposed to delocalize holes along the
Cu--O bonds by a soft bond stretching (LO) phonon (Fig. 
2, right), favoring hole exchange between two cages.
Notably the lattice dynamics of the
superconducting cuprates seems to distinguish itself from that of
other perovskites by an abrupt softening of this bond stretching phonon
at a wavevector $q\simeq 1/3$, signaling the onset of a charge order at
$T\leq 200$ K \cite{3}. We suggest low temperatures to 
favor ordering of the charge accumulating along the zig-zagging PSPS
at each third row of the Cu grid (Fig. 2, $left$).

\begin{figure}[th]
\centerline{\psfig{file=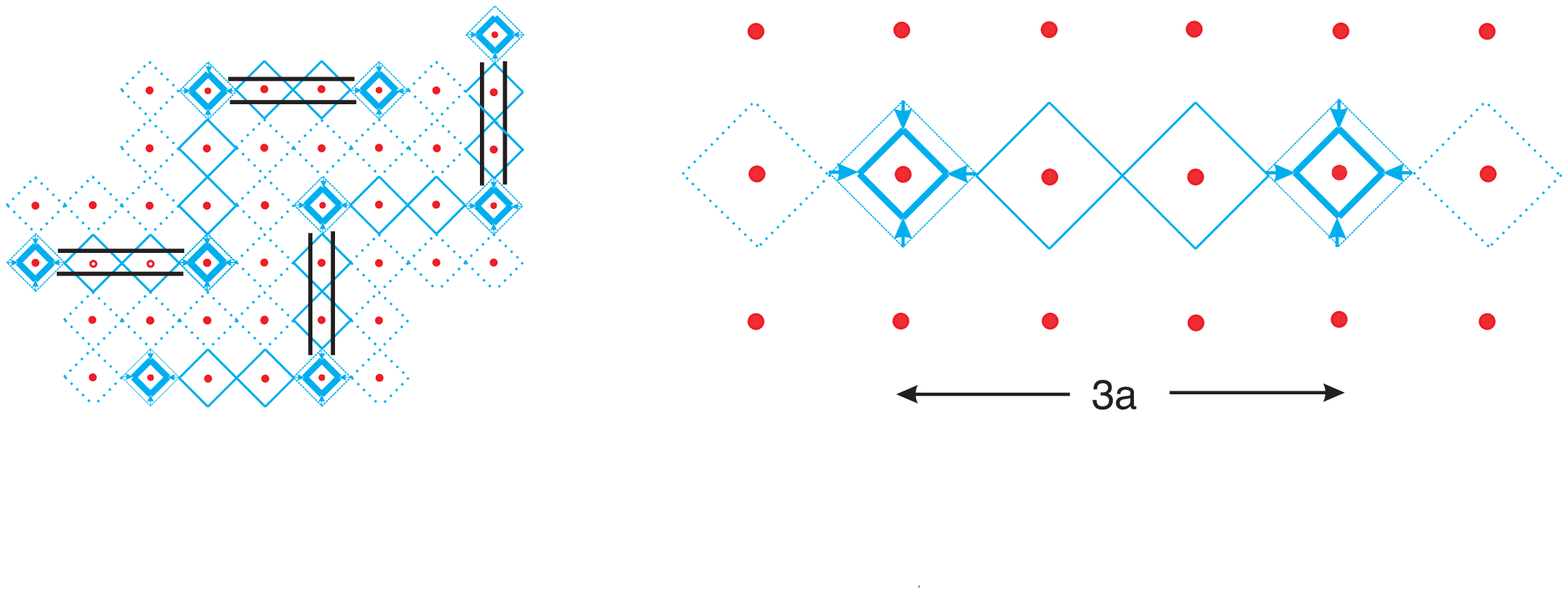,width=13cm,trim=2cm 9cm 1cm 5.2cm}}
\vspace*{8pt}
\caption{$Left$: Adiabatic snapshot of most closely packed PSPS. Thick 
parallel lines indicate double bonds zig-zagging along the Cu--O 
directions. $Right$: Scheme of the oxygen displacements of two SPS 
in the bond centered configuration.  The oxygen atoms in hole doped 
cages attract each other.}
\end{figure}


\begin{thebibliography}{0}

\bibitem{1} J. R\"ohler, Physica C {\bf 408-410}, 458 (2004), cond-mat/0304628.
    
\bibitem{2} J. R\"ohler, J. Supercond. {\bf 17}, 159 (2004), cond-mat/0307310.

\bibitem{3} H. Uchiyama, A.Q.R. Baron, S. Tsutsui, Y. Tanaka, W.-Z. Hu, 
A. Yamamoto, S. Tajima, Y. Endoh, Phys. Rev. Lett, {\bf 92}, 197005 
(2004), cond-mat/0403343.

\end{thebibliography}
\end{document}